\documentclass[
reprint,
superscriptaddress,
 amsmath,amssymb,
 aps,
prl
longbibliography
]{revtex4-2}

\usepackage{epsfig,amssymb,amsmath,graphicx,subfigure,hyperref}
\usepackage{tabularx}
\usepackage{float}
\usepackage{setspace}
\usepackage{color}
\usepackage{array} 
\newcommand{\PreserveBackslash}[1]{\let\temp=\\#1\let\\=\temp} \newcolumntype{C}[1]{>{\PreserveBackslash\centering}p{#1}} \newcolumntype{R}[1]{>{\PreserveBackslash\raggedleft}p{#1}} \newcolumntype{L}[1]{>{\PreserveBackslash\raggedright}p{#1}} 
\usepackage{graphicx}
\usepackage{dcolumn}
\usepackage{bm}
\usepackage{comment}
\usepackage{color}
\usepackage{soul}
\usepackage{ulem}

\begin{document}

\preprint{APS/123-QED}


\title{Programming active-molecule dynamics via intramolecular nonreciprocity}



\author{Ye Zhang}
\affiliation{Key Laboratory of Artificial Micro- and Nano-structures of Ministry of Education and School of Physics and Technology, Wuhan University, Wuhan 430072, China}

\author{Meng Xiao}
\affiliation{Key Laboratory of Artificial Micro- and Nano-structures of Ministry of Education and School of Physics and Technology, Wuhan University, Wuhan 430072, China}
\affiliation{Wuhan Institute of Quantum Technology, Wuhan 430206, China}

\author{Duanduan Wan}
\email[E-mail: ]{ddwan@whu.edu.cn}
\affiliation{Key Laboratory of Artificial Micro- and Nano-structures of Ministry of Education and School of Physics and Technology, Wuhan University, Wuhan 430072, China}
\date{\today}

\begin{abstract}
The dynamics of a self-propelled particle are typically hard-wired by its microscopic construction, limiting the range of behaviors accessible without redesigning the particle itself. Here we show that intramolecular nonreciprocity provides a minimal and versatile mechanism to overcome this constraint. We construct active molecules from short chains of two species of self-propelled particles whose propulsion directions are coupled nonreciprocally according to a prescribed internal sequence. At the single-molecule level, homogeneous sequences exhibit standard persistent random-walk dynamics, whereas heterogeneous sequences produce distinct trajectories inaccessible to either constituent species alone. At the collective level, using motility-induced phase separation (MIPS) as a representative example, we find that modifying the internal sequence shifts the MIPS onset by multiple orders of magnitude in propulsion strength, without altering particle-level interactions. These results demonstrate that intramolecular nonreciprocity among a small set of active components enables sequence-level programmability from single-molecule dynamics to emergent collective behavior, providing a minimal mechanism to encode and control active-matter dynamics across scales.
\end{abstract}

\maketitle

\textit{Introduction}---Active particles exhibit a wide range of nonequilibrium behaviors arising from the continuous injection of energy at the microscopic scale \cite{
Ramaswamy_2010_Mechanics,
Marchetti_2013_Hydrodynamicsa,
Shaebani_2020_Computationala,
shiExtremeSpontaneousDeformations2023a,
Mecke_2024_Emergent,
Gompper_2025_2025,
teVrugt_2025_Metareviewa}. However, the dynamical response of any given self-propelled particle is typically hard-wired by its microscopic construction, which prescribes how it translates, rotates, and responds to interactions with its surroundings 
\cite{
Vicsek_1995_Novel,
Toner_1995_Long-Range,
Caussin_2014_Emergent,
Banerjee_2017_Odd,
Liebchen_2017_Collective,
Doostmohammadi_2018_Activea,
Huang_2021_Circulara,
Zhang_2025_Self-propelled}. As a result, accessing new dynamical behaviors usually requires redesigning the particle itself. This limitation raises a basic question: how can one move beyond single-particle design rules to create simple, programmable structures that generate dynamical responses fundamentally inaccessible to individual active particles?

In active matter, a recently emerging route to enriching dynamics is the use of nonreciprocal interactions, in which the influence exerted by one particle on another is not symmetrically returned 
\cite{Banerjee_2017_Odd,Fruchart_2021_Non-reciprocal,Binysh_2022_Odd,Dinelli_2023_Non-reciprocitya,Veenstra_2025_Adaptivea}. Such interactions have been realized experimentally through chemical signaling \cite{Tucci_2024_Nonreciprocal,Soto_2014_Self-Assembly,Meredith_2020_Predator-prey}, hydrodynamic couplings \cite{Soni_2019_odd,Saha_2019_Pairing}, or externally imposed anisotropies \cite{
McNeill_2023_Acoustically,
Hanai_2025_Photoinduced,
Morrell_2025_Nonreciprocal}, and have been shown to produce interesting collective phenomena, for example, chiral motion \cite{Fruchart_2021_Non-reciprocal,Chen_2024_Emergent}, oscillations \cite{Saha_2020_Scalar,Frohoff-Hulsmann_2021_Suppression}, and nonreciprocal phase transitions \cite{Saha_2020_Scalar, Fruchart_2021_Non-reciprocal}. To date, however, nonreciprocity has been explored predominantly between distinct active units, shaping large-scale collective behavior in many-body active systems \cite{Saha_2020_Scalar,Fruchart_2021_Non-reciprocal,Dinelli_2023_Non-reciprocitya,Duan_2025_Phase}. By contrast, whether nonreciprocity can be embedded within an internally ordered, molecule-like structure and exploited as a programmable internal drive remains essentially unexplored. In particular, it remains unknown whether intramolecular nonreciprocity, combined with a tunable internal sequence of heterogeneous components, can generate distinct dynamical modes at the single-molecule level and systematically influence collective organization. This work shows that the internal sequence alone provides a programmable control knob linking single-molecule dynamics to emergent collective behavior.

Here we develop a minimal framework in which the internal sequence serves as the primary design variable. The central point is that programmability is achieved at the sequence level, without altering any particle level interactions. We construct active molecules from short chains of two species of self-propelled particles connected in prescribed sequences and coupled through nonreciprocal orientation interactions. Like-type neighbors couple reciprocally, whereas cross-type neighbors couple nonreciprocally: one species tends to align to the other, while the reverse coupling favors a different orientational response, generating sequence-dependent internal drives. This architecture yields qualitatively new dynamics on two levels. At the single-molecule level, homogeneous sequences display standard persistent random-walk motion, whereas heterogeneous sequences produce distinct, sequence-specific trajectories that are inaccessible to either constituent species alone. At the collective level, we focus on motility-induced phase separation (MIPS) \cite{Buttinoni_2013_Dynamical, Redner_2013_Structure, Elgeti_2015_Physics, Bechinger_2016_Activeb, Digregorio_2018_Full, Omar_2021_Phasea, maDynamicalClusteringInterrupts2022} as an example and show that altering the internal sequence shifts the MIPS onset by multiple orders of magnitude in propulsion strength, without changing particle-level interactions. Together, these results demonstrate that intramolecular nonreciprocity, when combined with sequence programmability, provides a simple and versatile strategy for sequence-encoded control of both single-molecule dynamics and emergent collective behavior in active matter. This establishes intramolecular nonreciprocity with tunable internal sequences as a general design principle for programmable active matter.

\begin{figure}
\centering
\includegraphics[width =1\linewidth]{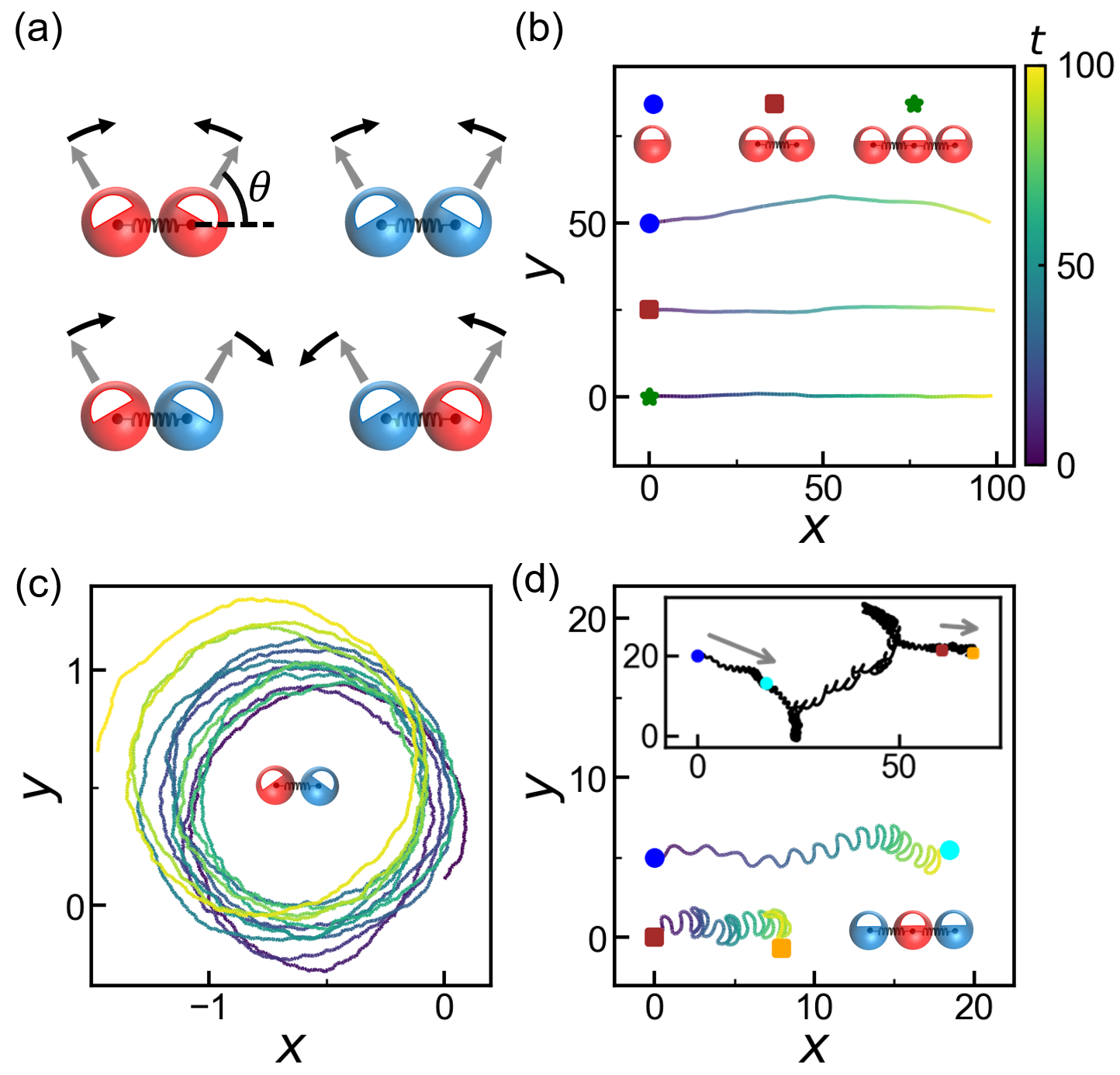}   
\caption{(a) Illustration of the nonreciprocal coupling between the propulsion orientations of two bonded particles at $g=-1$. Red and light blue denote $A$ and $B$ particles. White semicircles and gray arrows indicate the instantaneous self-propulsion directions; for each particle, the orientation angle $\theta$ is measured relative to a fixed reference direction (dashed line).
(b–d) Representative center-of-mass trajectories of molecular sequences at $g=-1$, shown over a time window of duration $\Delta t = 100$. Trajectories are shifted in time to start at $t=0$ and translated so that the initial position is at $x=0$.
(b) Trajectories of the [A], [A–A], and [A–A–A] sequences. For clarity, trajectories are vertically shifted to different y positions.
(c) Trajectory of the [A-B] sequence.
(d) Two representative trajectories of the [B–A–B] sequence. The inset shows a long-time trajectory, with the starting and ending points of the two representative segments indicated by different symbols.}
\label{fig:model}
\end{figure}

\textit{Model}---We consider a simple model in which nonreciprocal interactions act only within a molecule. A molecule is a linear chain of $m$ self-propelled particles of two species, $A$ and $B$, ordered in a prescribed sequence, with nonreciprocal alignment interactions between bonded neighbors. See Fig.~1(a) for illustrative $m=2$ sequences. The overdamped Langevin equations governing the dynamics of particle $i$ read \cite{Digregorio_2018_Full,Redner_2013_Structure}
\begin{equation}
\begin{aligned}
\dot{\mathbf{r}}_i
&= \frac{1}{\gamma} \!\left[ F_0\,\mathbf{n}_i
- \nabla_i \!\left( \sum_{j\in\mathcal B(i)} U_{\mathrm{spr}}(r_{ij}) + \sum_{k (\neq i)} U_{\mathrm{WCA}}(r_{ik}) \right) \right] \\
&\quad + \sqrt{2 D_\text{t}}\,\bm{\eta}^t_i ,
\end{aligned}
\label{eq:trans}
\end{equation}
\begin{equation}
\dot{\theta}_{i}
= \sqrt{2D_{\mathrm r}}\,\eta_{i}^{\mathrm r}
+ \sum_{j\in\mathcal B(i)} M_{ij}\,\sin(\theta_{j}-\theta_{i}),
\label{eq:angle}
\end{equation}
where $\mathbf{r}_i$ and $\mathbf{n}_i=(\cos\theta_i,\sin\theta_i)$ denote the position and propulsion direction of particle $i$, respectively. Here $\theta_i$ is the instantaneous propulsion orientation of particle $i$, measured relative to a fixed reference direction [Fig.~\ref{fig:model}(a)]. The translational and rotational noises, $\bm{\eta}^t_i$ and $\eta_i^{\mathrm r}$, are Gaussian white noises with zero mean and unit variance. Bonded particles within the same molecule interact via a harmonic spring potential $U_{\mathrm{spr}}(r)=k(r-r_0)^2/2$; in Eq.~(\ref{eq:trans}), $\mathcal{B}(i)$ denotes the set of particles bonded to $i$. Excluded-volume interactions between all particle pairs, both within and between molecules, are modeled by the Weeks–Chandler–Andersen (WCA) potential $U_{\mathrm{WCA}}(r)=4\epsilon \left[(\sigma/r)^{12}-(\sigma/r)^6\right]+\epsilon$ for $r<2^{1/6}\sigma$, and $U_{\mathrm{WCA}}(r)=0$ otherwise. Nonreciprocal orientational coupling is encoded in the matrix $M_{ij}=J_{s_i s_j}$, where $s_i\in\{A,B\}$ denotes the particle species and
\begin{equation}
\begin{pmatrix}
J_{AA} & J_{AB}\\
J_{BA} & J_{BB}
\end{pmatrix}
=
\begin{pmatrix}
1 & 1\\
g & 1
\end{pmatrix}.
\label{eq:J_matrix}
\end{equation}
For $g\neq 1$, the coupling between $A$ and $B$ particles becomes asymmetric, breaking action–reaction symmetry. The translational and rotational diffusion coefficients are related by
$D_{\mathrm t}=D_{\mathrm r}\sigma^2/3$, with $D_{\mathrm t}=k_B T/\gamma$.
In our simulations, we set $\sigma=\gamma=1$, fix $J_{AA}=1$, and measure time in units of $J_{AA}^{-1}$. We choose the spring stiffness and the WCA energy scale to increase linearly with $F_0$. Unless otherwise specified, we set $D_{\mathrm r}=10^{-3}$ and $F_0=1$. Numerical integration is performed using HOOMD-blue’s overdamped Brownian module \cite{Anderson_2020_HOOMDblue}; see Supplemental Material (SM, Sec.~II) for further simulation details.

\textit{Results}---Figure \ref{fig:model}(a) illustrates the nonreciprocal orientational coupling between two bonded particles for $g=-1$. For identical species, the interaction drives alignment of the propulsion directions toward a common orientation. For unlike species, the orientations instead co-rotate with a constant phase lag. Figures~\ref{fig:model}(b)-\ref{fig:model}(d) show representative center-of-mass trajectories. Homogeneous molecules $[\mathrm{A}^m]$ exhibit progressively smoother trajectories as $m$ increases from 1 to 3 [Fig.~\ref{fig:model}(b)], consistent with averaging of directional fluctuations in longer chains. Heterogeneous molecules behave qualitatively differently: an [A-B] dimer exhibits pronounced circling [Fig.~\ref{fig:model}(c)], while longer heterogeneous sequences combine features of both homogeneous chains and the dimer. For example, the [B–A–B] trimer displays a wriggling forward motion [Fig.~\ref{fig:model}(d)]. Two segments from the same long trajectory (inset) show similar forward drift but different wriggling frequencies, reflecting noise-induced variations in the interparticle phase differences that modulate the wriggling pattern. See also Supplemental Movies S1–S2 for the trajectories. We also verified that the alternating trimer [A--B--A] exhibits a similar wriggling mode and admits an analogous analysis.

\begin{figure}[t]
\centering
\includegraphics[width=1\linewidth]{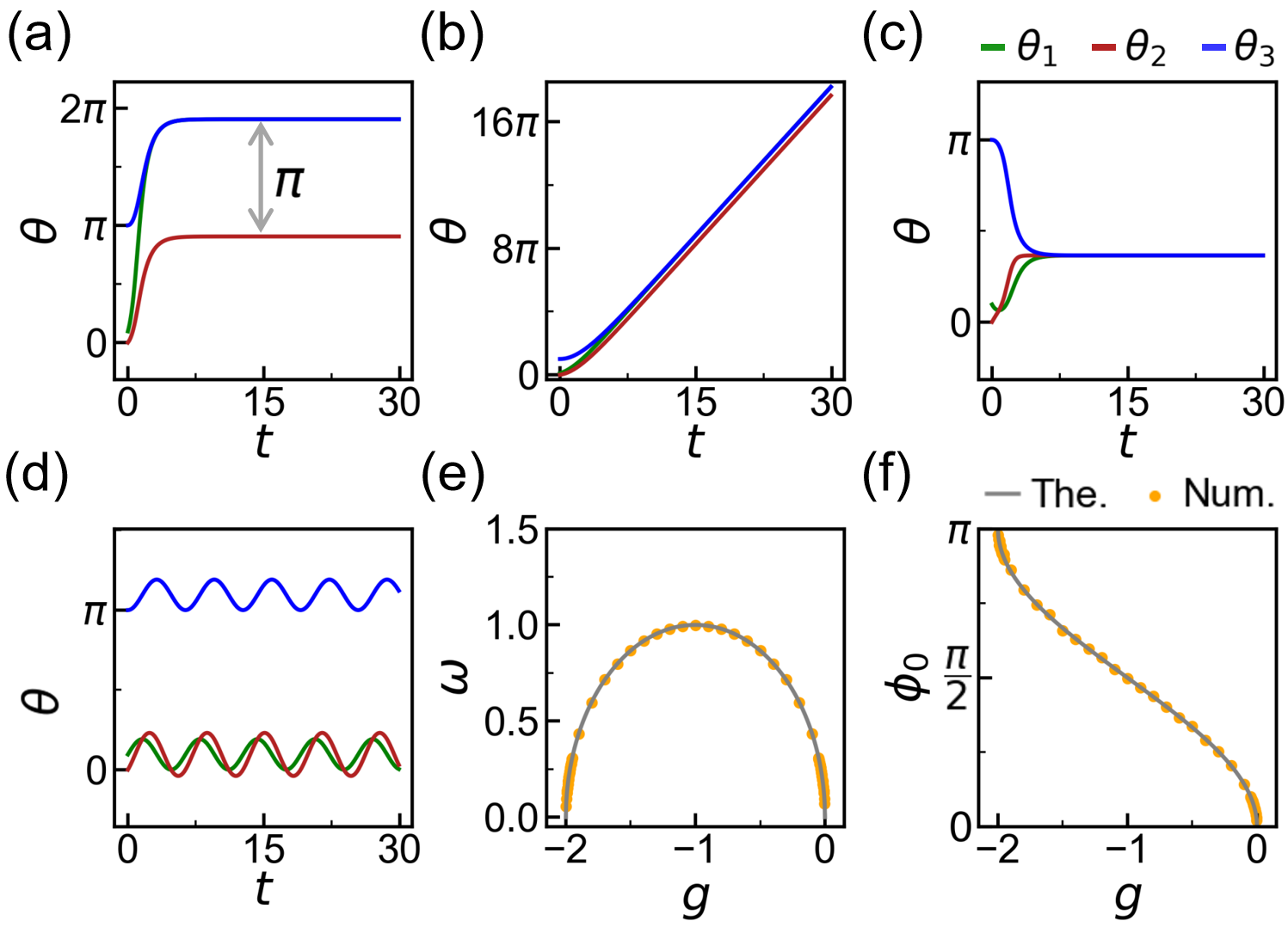}
\caption{Angular dynamics of a [B–A–B] molecule.
(a–d) Time evolution of the orientation angles $\theta_i(t)$ for an initial condition $(\theta_1,\theta_2,\theta_3)=(0.3,0,\pi)$, shown for (a) $g=-3$, (b) $g=-2$, (c) $g=1$, and (d) $g=-1$.
(e,f) In the oscillatory regime $-2<g<0$, oscillation frequency $\omega$ (e) and phase offset $\phi_0$ (f), defined in Eq.~(\ref{eq:nonreciprocal}), as functions of $g$.
}
\label{fig:changeG}
\end{figure}

To isolate circling and wriggling motions arising solely from nonreciprocal alignment, we first consider the deterministic limit $D_{\mathrm t}=D_{\mathrm r}=0$. The circling dynamics of the [A–B] dimer are discussed in SM, Sec.~III; here we focus on the [B–A–B] trimer, which already captures the essential features of longer sequences. The orientational dynamics depend strongly on the nonreciprocity parameter $g$ and fall into several distinct regimes (see SM, Sec.~III for a complete classification). Figures~\ref{fig:changeG}(a–d) show representative trajectories starting from $(\theta_1,\theta_2,\theta_3)=(0.3,0,\pi)$. For strong nonreciprocity, $g<-2$ [e.g., $g=-3$ in Fig.~\ref{fig:changeG}(a)], the molecule relaxes to a static phase-locked state in which each $B$ particle anti-aligns with the central $A$, yielding fixed relative angles $\theta_1-\theta_2=\theta_3-\theta_2=\pi$. At the special point $g=-2$ [Fig.~\ref{fig:changeG}(b)], the relative angles again phase-lock, but the absolute orientations do not become stationary; instead, all three particles co-rotate with a constant angular velocity $\Omega$, $\theta_i(t)=\Omega t+\phi_i$, forming a uniformly rotating, phase-locked state. For $g>0$ [e.g., $g=1$ in Fig.~\ref{fig:changeG}(c)], reciprocal alignment dominates and the molecule relaxes to a fully aligned stationary state, $\theta_1=\theta_2=\theta_3$. Between these limits, in the intermediate regime $-2<g<0$ [e.g., $g=-1$ in Fig.~\ref{fig:changeG}(d)], the relative angles undergo persistent oscillations, producing the wriggling motion of the molecule. Defining $x\equiv\theta_1-\theta_2$ and $y\equiv\theta_3-\theta_2$, linearization about either stable fixed point $(x^*, y^*)=(0,\pi)$ or $(\pi,0)$ yields
\begin{equation}
\begin{pmatrix}
x \\[4pt]
y
\end{pmatrix}
=
\begin{pmatrix}
x^* \\[4pt]
y^*
\end{pmatrix}
+
a
\begin{pmatrix}
\cos(\omega t) \\[4pt]
\cos(\omega t+\phi_0)
\end{pmatrix},
\label{eq:nonreciprocal}
\end{equation}
with $\omega=\sqrt{-g(g+2)}$ and $\phi_0=\arccos(1+g)$ (see SM, Sec.~III for derivations). Figures~\ref{fig:changeG}(e,f) confirm these predictions throughout $-2<g<0$, showing excellent agreement between simulations and theory.

\begin{figure}[t]
    \centering
    \includegraphics[width=1\linewidth]{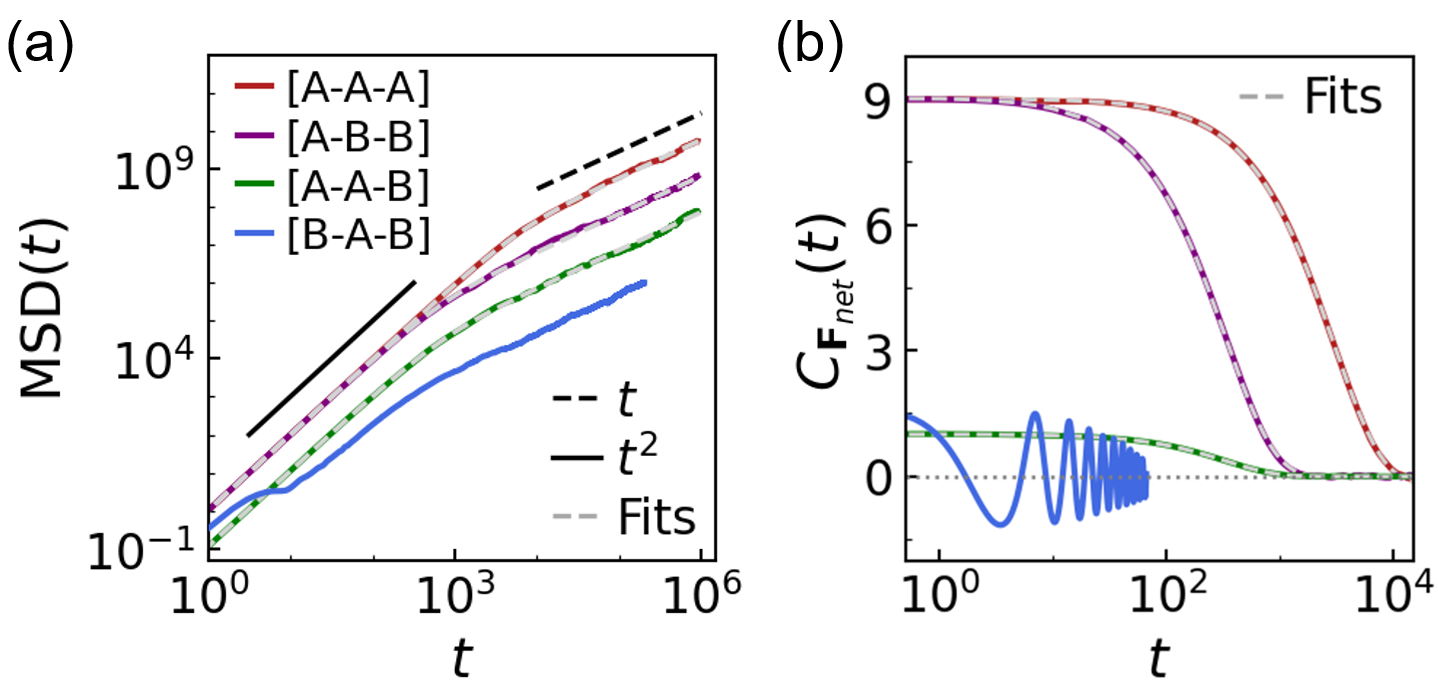}
\caption{
(a) Mean-squared displacement (MSD) for four three-particle molecular sequences. Gray dashed curves are given by Eq.~(\ref{eq:msd_theory}), using the effective parameters extracted from the autocorrelation in (b). For [B–A–B], data are shown up to $t=2\times10^{5}$ because resolving its rapid orientational fluctuations requires a smaller integration time step (see SM, Sec.~V). (b) Autocorrelation of the net propulsion force, $C_{\mathbf{F}_{\mathrm{net}}}(t)$ [Eq.~(\ref{eq:corre})], for the same sequences. Gray dashed lines are exponential fits; the gray dotted line indicates $C_{\mathbf{F}_{\mathrm{net}}}(t)=0$.
}
\label{fig:fig3auto}
\end{figure}

For active-matter systems such as Brownian particles, fluctuations due to thermal noise are unavoidable. We therefore turn to the behavior of sequence-encoded molecules at finite noise and examine how the internal sequence controls both single-molecule dynamics and collective behavior in the presence of nonreciprocal couplings. Unless otherwise specified, we focus on the representative case $g=-1$; varying $g$ provides continuous control over the dynamics, and results for other values of $g$ are summarized in SM, Sec.~IV.
We characterize molecular motion via the mean-squared displacement (MSD) of the center of mass,
\begin{equation}
\mathrm{MSD}(t)=
\Big\langle
\big|\mathbf{R}(t_{0}+t)-\mathbf{R}(t_{0})\big|^{2}
\Big\rangle_{t_{0}},
\label{eq:msd_def}
\end{equation}
where $\mathbf{R}(t)=\frac{1}{3}\sum_{i=1}^{3}\mathbf{r}_i(t)$ is the molecular center-of-mass position and $\langle\cdot\rangle_{t_{0}}$ denotes an average over reference times $t_{0}$ (and independent trajectories; see SM, Sec.~V for numerical details). Figure~\ref{fig:fig3auto}(a) shows $\mathrm{MSD}(t)$ for four molecular sequences. For [A–A–A], [A–B–B], and [A–A–B], the MSD exhibits the familiar two-regime behavior of active Brownian motion: an initial ballistic regime with $\mathrm{MSD}(t)\sim t^{2}$ (solid guide), followed by a long-time diffusive regime with $\mathrm{MSD}(t)\sim t$ (dashed guide). The crossover time depends strongly on the molecular code: for [A–A–A] the ballistic-to-diffusive crossover extends to $t\approx10^{4}$, whereas for [A–B–B] and [A–A–B] it occurs earlier, at $t\approx10^{3}$. The MSD curves for [A–B–B] and [A–A–B] have similar shapes, with [A–B–B] shifted upward by an approximately constant offset. In contrast, the wriggling sequence [B–A–B] deviates from a clean ballistic–diffusive two-regime form and exhibits a more intricate crossover. 

To rationalize these behaviors, we analyze temporal correlations of the net propulsion force via its autocorrelation function,
\begin{equation}
C_{\mathbf{F}_{\mathrm{net}}}(t)
=
\left\langle 
  \mathbf{F}_{\mathrm{net}}(t_{0}+t)
  \cdot
  \mathbf{F}_{\mathrm{net}}(t_{0})
\right\rangle_{t_{0}},
\label{eq:corre}
\end{equation}
where
\(
  \mathbf{F}_{\mathrm{net}}(t)
  =
  \sum_{i=1}^{3} F_{0} \mathbf{n}_i(t)
\)
is the instantaneous net active propulsion force of a molecule (see SM, Sec.~V for numerical details). Figure~\ref{fig:fig3auto}(b) shows $C_{\mathbf{F}_{\mathrm{net}}}(t)$ for the four sequences. For [A–A–A], [A–B–B], and [A–A–B], the data are well described by an exponential form,
$C_{\mathbf{F}_{\mathrm{net}}}(t)\simeq F_{\mathrm{eff}}^{2}\exp(-D_{\mathrm{eff}}t)$
(see SM, Sec.~V). The corresponding fits (dashed lines) yield $F_{\mathrm{eff}}=3F_{0}$ and $D_{\mathrm{eff}}=D_{\mathrm r}/3$ for [A–A–A], $F_{\mathrm{eff}}=3F_{0}$ and $D_{\mathrm{eff}}=3D_{\mathrm r}$ for [A–B–B], and $F_{\mathrm{eff}}=F_{0}$ and $D_{\mathrm{eff}}=3D_{\mathrm r}$ for [A–A–B]. Using these parameters, one obtains the analytical MSD (see SM, Sec.~V),
\begin{equation}
\mathrm{MSD}(t)
=
\frac{4D_{\mathrm t}}{3}\,t
+\frac{2}{9\gamma^2}F_{\mathrm{eff}}^2
\left[
\frac{t}{D_{\mathrm{eff}}}
-\frac{1-e^{-D_{\mathrm{eff}}t}}{D_{\mathrm{eff}}^{2}}
\right],
\label{eq:msd_theory}
\end{equation}
shown as dashed curves in Fig.~\ref{fig:fig3auto}(a), in good agreement with the simulation results. By contrast, for the wriggling sequence [B–A–B], $C_{\mathbf{F}_{\mathrm{net}}}(t)$ exhibits pronounced oscillations about zero, reflecting the underlying oscillatory orientational dynamics.

These results show that even simple molecular sequences generate markedly different force autocorrelations and hence distinct levels of directional memory: a slower decay of $C_{\mathbf{F}_{\mathrm{net}}}(t)$ implies a longer-lived net propulsion direction and a larger persistence length. This sequence-dependent persistence provides a direct route to shifting collective thresholds. To demonstrate how the same molecular code controls collective behavior, we focus on MIPS, where crowding-induced motility reduction yields dense--dilute coexistence in the absence of attractive interactions~\cite{Buttinoni_2013_Dynamical,Redner_2013_Structure,Elgeti_2015_Physics,Bechinger_2016_Activeb,Digregorio_2018_Full,Omar_2021_Phasea}. To quantify phase separation, we use the average fraction of molecules in the largest cluster, $\bar{\alpha}$, as an order parameter:
\begin{equation}
\bar{\alpha}=\big\langle N_{\mathrm{lc}}(t)/N\big\rangle,
\label{eq:alpha}
\end{equation}
where $N$ is the total number of molecules and $N_{\mathrm{lc}}(t)$ is the size of the largest cluster at time $t$. The average is taken in the quasi-steady state and over independent configurations (see SM, Sec.~VI).

\begin{figure}[t]
    \centering
    \includegraphics[width=1.0\linewidth]{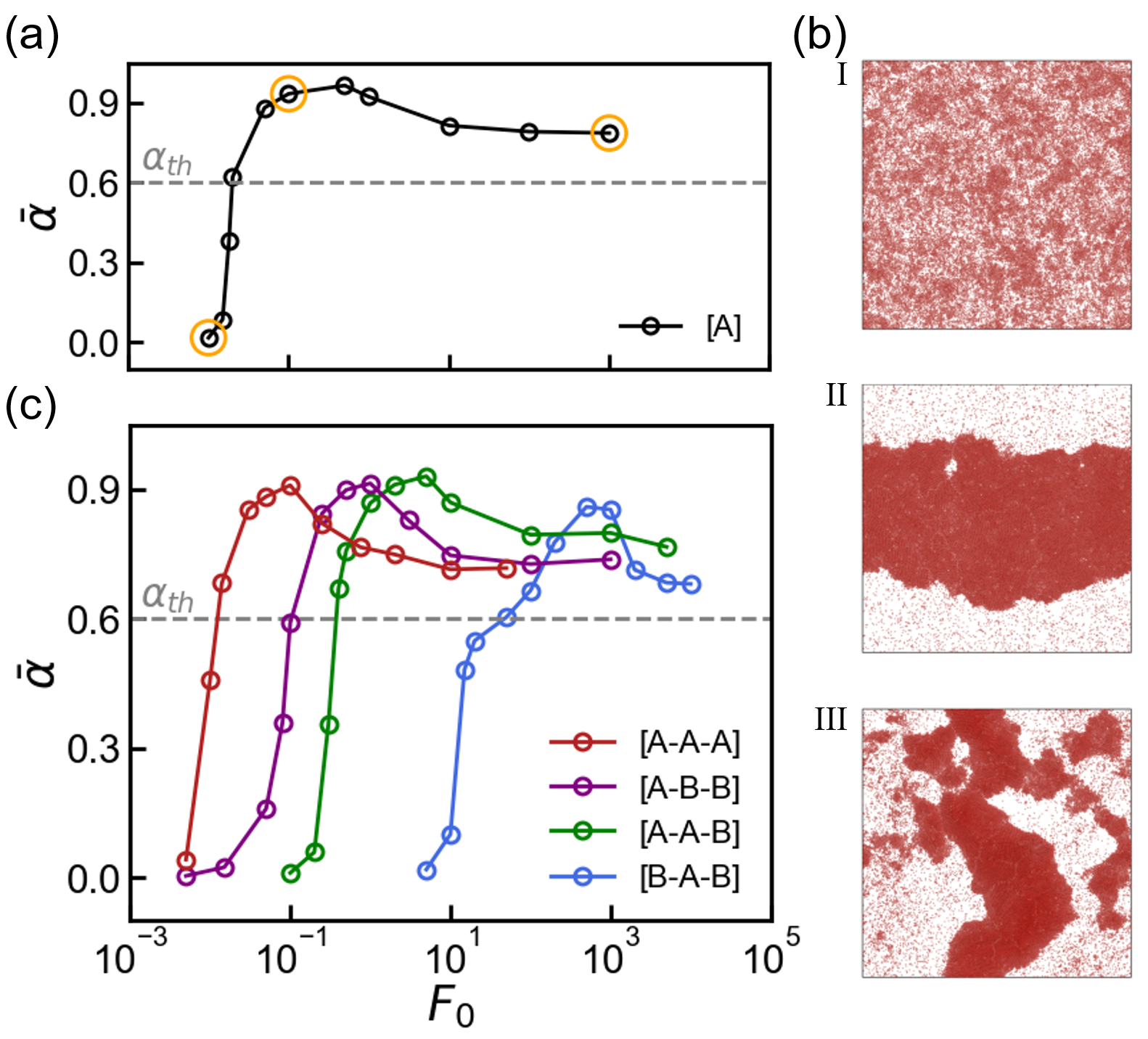}
    \caption{ 
(a) Average fraction of molecules in the largest cluster, $\bar{\alpha}$ [Eq.~(\ref{eq:alpha})], as a function of propulsion strength $F_{0}$ for monomeric [A] molecules ($N=43{,}200$). The gray dashed line marks the threshold $\alpha_{\mathrm{th}}=0.6$.
(b) Representative snapshots corresponding to the orange open circles in (a).
(c) $\bar{\alpha}$ versus $F_{0}$ for four trimer sequences ($N=14{,}400$).
    }
\label{fig:fig3mips}
\end{figure}

Figure~\ref{fig:fig3mips}(a) shows $\bar{\alpha}$ versus propulsion strength $F_{0}$ for monomeric $A$ molecules. As $F_{0}$ increases, $\bar{\alpha}$ rises sharply from near zero and reaches a maximum of about $0.9$, and then decreases gradually at larger $F_{0}$. Representative snapshots are shown in Fig.~\ref{fig:fig3mips}(b): the system is nearly homogeneous at small $F_{0}$ [Fig.~\ref{fig:fig3mips}(b${\mathrm{I}}$)], becomes strongly phase separated at intermediate $F_{0}$ [Fig.~\ref{fig:fig3mips}(b${\mathrm{II}}$)], and exhibits fragmentation of large clusters at large $F_{0}$ [Fig.~\ref{fig:fig3mips}(b${\mathrm{III}}$)], leading to a reduction of $\bar{\alpha}$. Similar trends are observed for trimers, as shown in Fig.~\ref{fig:fig3mips}(c). We define the onset of MIPS operationally by a threshold $\alpha_{\mathrm{th}}=0.6$ and denote by $F_{0,\mathrm{th}}$ the value of $F_{0}$ at which $\bar{\alpha}$ first reaches $\alpha_{\mathrm{th}}$. Remarkably, $F_{0,\mathrm{th}}$ varies by orders of magnitude across the four molecular codes, with interparticle interactions held fixed:
\begin{equation}
F_{0,\mathrm{th}}^{\mathrm{[A\!-\!A\!-\!A]}} 
< F_{0,\mathrm{th}}^{\mathrm{[A\!-\!B\!-\!B]}} 
< F_{0,\mathrm{th}}^{\mathrm{[A\!-\!A\!-\!B]}} 
< F_{0,\mathrm{th}}^{\mathrm{[B\!-\!A\!-\!B]}}.
\label{eq:Fcritical}
\end{equation}
This ordering is consistent with the single-molecule dynamics. The [A–A–A] code yields the longest persistence, with $F_{\mathrm{eff}}=3F_{0}$ and the smallest effective rotational diffusion $D_{\mathrm{eff}}=D_{r}/3$, and thus exhibits the smallest $F_{0,\mathrm{th}}$. The [A–B–B] code has the same $F_{\mathrm{eff}}=3F_{0}$ but a larger $D_{\mathrm{eff}}=3D_{r}$, requiring a higher $F_{0,\mathrm{th}}$. The [A–A–B] code has a reduced effective propulsion, $F_{\mathrm{eff}}=F_{0}$ (with $D_{\mathrm{eff}}=3D_{r}$), and therefore an even larger $F_{0,\mathrm{th}}$. Finally, the wriggling [B–A–B] code produces rapid decorrelation of the propulsion direction, strongly suppressing persistence; consequently, substantially larger $F_{0}$ is needed to induce MIPS, yielding the largest $F_{0,\mathrm{th}}$.

In conclusion, we studied active molecules assembled from short chains of two self-propelled species whose propulsion directions are coupled nonreciprocally according to a prescribed internal sequence. At the single-molecule level, homogeneous sequences exhibit standard persistent random-walk motion, whereas heterogeneous sequences generate distinct, sequence-specific trajectories that are inaccessible to either constituent species alone. At the collective level, we show that sequence reprogramming shifts the onset of MIPS by orders of magnitude in propulsion strength, while keeping particle-level interactions unchanged. More broadly, the same coding strategy can be extended to other collective phenomena in active matter. A natural next step is to incorporate additional intramolecular degrees of freedom (e.g., internal dipoles) that mediate alignment and angular-momentum exchange between molecules, and to explore the resulting emergent phases. Overall, our results establish sequence-encoded intramolecular nonreciprocity as a minimal design principle for programming active-matter dynamics across scales, from single-molecule motion to collective organization.

\begin{acknowledgments}
This work was supported by the National Natural Science Foundation of China (Grants No.~12274330, No.~12334015, and No.~12274332) and the Young Top-Notch Talent Cultivation Program of Hubei Province.
D.W. acknowledges the ``Xiaomi Young Scholar Program" at Wuhan University. 

\end{acknowledgments}

\bibliography{main}

\end{document}


\title{\large Supplemental Materials for:\\
“Programmable active phases via nonreciprocal molecular coding”}

\author{Ye Zhang}
\affiliation{Key Laboratory of Artificial Micro- and Nano-structures of Ministry of Education and School of Physics and Technology, Wuhan University, Wuhan 430072, China}

\author{Meng Xiao}
\affiliation{Key Laboratory of Artificial Micro- and Nano-structures of Ministry of Education and School of Physics and Technology, Wuhan University, Wuhan 430072, China}
\affiliation{Wuhan Institute of Quantum Technology, Wuhan 430206, China}

\author{Duanduan Wan}
\email{ddwan@whu.edu.cn}
\affiliation{Key Laboratory of Artificial Micro- and Nano-structures of Ministry of Education and School of Physics and Technology, Wuhan University, Wuhan 430072, China}
\maketitle

\tableofcontents
\newpage

\supsection{Captions for Movies S1 to S3}

Supplemental Movies S1--S3 visualize representative molecular trajectories. 

\textbf{Movie S1 (Movie\_S1\_single\_molecule\_trajectories.mp4):}
Single-molecule trajectories for the sequences \textbf{[A]}, \textbf{[A--A]}, \textbf{[A--A--A]}, \textbf{[A--B--B]}, and \textbf{[A--A--B]}. Trajectories are color-coded by time. All parameters match those in Fig.~1 of the main text.

\textbf{Movie S2 (Movie\_S2\_circling\_and\_wriggling.mp4):}
Single-molecule trajectories for the sequences \textbf{[A--B]} and \textbf{[B--A--B]}. Trajectories are color-coded by time. All parameters match those in Fig.~1 of the main text.

\textbf{Movie S3 (Movie\_S3\_MIPS\_AAA\_vs\_BAB.mp4):}
Collective dynamics illustrating motility-induced phase separation (MIPS) in a system of $N=14{,}400$ molecules with sequences \textbf{[A--A--A]} and \textbf{[B--A--B]} at $F_0=10$. All parameters are the same as in Fig.~4 of the main text. For clarity, the trajectory of one representative molecule of each sequence is highlighted and color-coded by time. In the rendering, $A$ particles are colored red and $B$ particles blue; at the current movie resolution these colors may appear gray.

\supsection{Simulation details and model parameters}

All simulations are performed using overdamped Brownian dynamics implemented in HOOMD-blue \cite{Anderson_2020_HOOMDblue}. The translational and rotational noises, $\bm{\eta}^t_i(t)$ and $\eta^r_i(t)$, are Gaussian white noises with zero mean and unit variance:
$\langle \bm{\eta}^t_i(t)\rangle=\mathbf{0}$,
$\langle \bm{\eta}^t_i(t)\bm{\eta}^t_j(t')\rangle=\delta_{ij}\delta(t-t')\,\mathbf{1}$,
$\langle \eta^r_i(t)\rangle=0$, and
$\langle \eta^r_i(t)\eta^r_j(t')\rangle=\delta_{ij}\delta(t-t')$.

Lengths are measured in units of $\sigma$, with $\sigma=1$, and the friction coefficient is set to $\gamma=1$. We fix the orientational coupling strength $J_{AA}=1$ and use $J_{AA}^{-1}$ as the unit of time. Except in the noise-free limit (where $D_{\mathrm{r}}=0$), we set the rotational diffusion coefficient to $D_{\mathrm{r}}=10^{-3}$, which determines the translational diffusion coefficient via $D_{\mathrm{t}}=D_{\mathrm{r}}\sigma^2/3$. With $\gamma=1$, this choice also fixes the thermal energy scale through $D_{\mathrm{t}}=k_BT/\gamma$.

Bonded particles within the same molecule interact via a harmonic spring potential with equilibrium bond length $r_0=2^{1/6}$ and stiffness $k=60F_0$, ensuring stiff bonds that confine particles within each molecule. Excluded-volume interactions between all particle pairs are modeled by the Weeks--Chandler--Andersen (WCA) potential with energy scale $\epsilon=0.5F_0$. Scaling both $k$ and $\epsilon$ linearly with the activity $F_0$ avoids effective softening at large propulsion forces and keeps the relative interaction stiffness approximately constant \cite{Su_2023_Motility-induced}. The integration time step is chosen as $\mathrm{d}t=3\times10^{-4}/F_0$, which provides numerical stability across the full range of activities considered.

In the MIPS simulations, $F_0$ is varied while all other parameters are kept fixed; in all other simulations we set $F_0=1$. For the single-molecule trajectories shown in Fig.~1 of the main text, each run is carried out for $10^{4}$ time units. The first $10^{3}$ time units are discarded as equilibration, and the remaining trajectory is used for analysis.

\supsection{Single-molecule orientational dynamics in the noise-free limit}
\label{sec:no_noise}

\supsubsection{Two-particle [A--B] molecule}

We first consider a two-particle molecule with one $A$ and one $B$ particle to illustrate the basic alignment mechanism, and then extend the analysis to three-particle molecules. Let $\theta_1(t)$ and $\theta_2(t)$ denote the propulsion orientations of particles $A$ and $B$, respectively. In the deterministic limit ($D_{\mathrm{t}}=D_{\mathrm{r}}=0$), Eq.~(2) of the main text gives
\begin{equation}
\dot{\theta}_{1}=\sin(\theta_2-\theta_1), \qquad
\dot{\theta}_{2}=g\,\sin(\theta_1-\theta_2).
\end{equation}
Introducing the relative angle $x\equiv\theta_1-\theta_2$, the dynamics reduce to
\begin{equation}
\dot{x}=-(1+g)\sin x .
\end{equation}
The fixed points are $x^*=0$ (alignment) and $x^*=\pi$ (anti-alignment). Linearizing about a fixed point, $\delta x\equiv x-x^*$, yields
\begin{equation}
\dot{\delta x}=-(1+g)\cos x^{*}\,\delta x .
\end{equation}
Therefore, the aligned state is stable for $g>-1$, while the anti-aligned state is stable for $g<-1$. At the transition point $g=-1$, the relative angle is conserved, $x(t)\equiv x_0$, and the two particles co-rotate with the same angular velocity,
$\dot{\theta}_1=\dot{\theta}_2=-\sin x_0$.
In summary, $g$ selects the long-time dynamics: alignment ($x\to0$) for $g>-1$, anti-alignment ($x\to\pi$) for $g<-1$, and rigid co-rotation with a fixed phase lag at $g=-1$.

\supsubsection{Three-particle [B--A--B] molecule}

\begin{figure}
    \centering
    \includegraphics[width=0.70\linewidth]{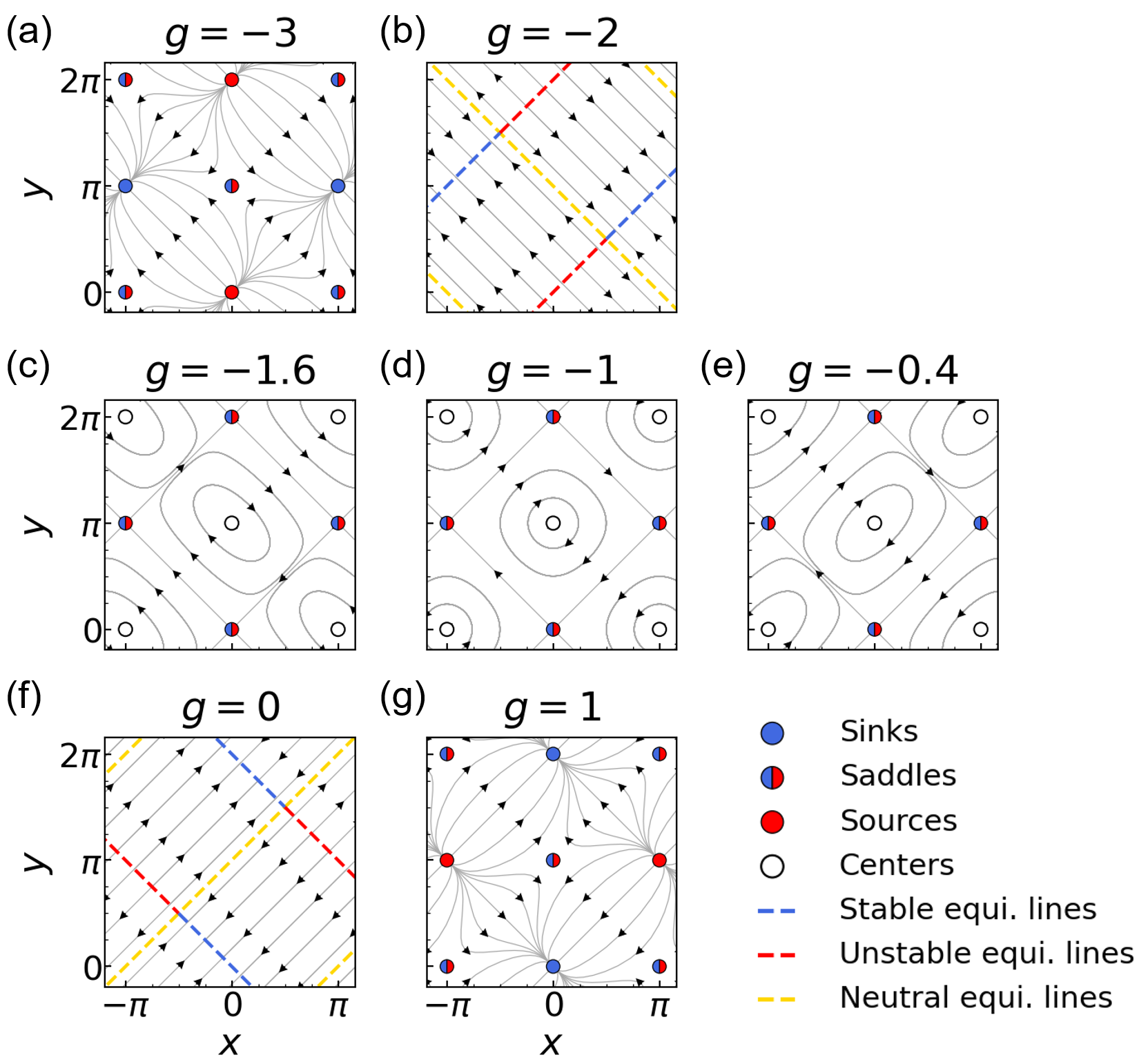}
\caption{
Flow in the $(x,y)$ plane for the [B--A--B] molecule [Eq.~\eqref{eq:BAB_xy}] at representative values of $g$. Gray lines are trajectories, and arrows indicate the direction of motion.
}
\label{fig:point}
\end{figure}

We next consider a three-particle molecule in the [B--A--B] sequence. In the deterministic limit, the orientational dynamics read
\begin{equation}
\begin{cases}
\dot{\theta}_{1} = g\,\sin(\theta_{2}-\theta_{1}),\\[2pt]
\dot{\theta}_{2} = \sin(\theta_{1}-\theta_{2})+\sin(\theta_{3}-\theta_{2}),\\[2pt]
\dot{\theta}_{3} = g\,\sin(\theta_{2}-\theta_{3}).
\end{cases}
\end{equation}
Defining the relative angles
\begin{equation}
x \equiv \theta_1-\theta_2,\qquad y \equiv \theta_3-\theta_2,
\end{equation}
one obtains
\begin{equation}
\dot{x}=-(1+g)\sin x-\sin y,\qquad
\dot{y}=-\sin x-(1+g)\sin y .
\label{eq:BAB_xy}
\end{equation}
Since $x$ and $y$ are $2\pi$-periodic, we restrict to a $2\pi$-periodic domain. Within this domain, there are four fixed points,
\begin{equation}
(x^{*},y^{*})\in\{(0,0),(\pi,\pi),(0,\pi),(\pi,0)\}.
\end{equation}
Linearizing about $(x^*,y^*)$ with $\delta x=x-x^*$ and $\delta y=y-y^*$ gives
\begin{equation}
\frac{\mathrm{d}}{\mathrm{d}t}
\begin{pmatrix}
\delta x \\[2pt]
\delta y
\end{pmatrix}
=
\mathcal{J}
\begin{pmatrix}
\delta x \\[2pt]
\delta y
\end{pmatrix},
\qquad
\mathcal{J}=
\begin{pmatrix}
-(1+g)\cos x^{*} & -\cos y^{*}\\[2pt]
-\cos x^{*} & -(1+g)\cos y^{*}
\end{pmatrix}.
\end{equation}
The corresponding eigenvalues are
\begin{equation}
\begin{aligned}
(0,0):\;&\lambda_1=-g,\quad \lambda_2=-(g+2),\\
(\pi,\pi):\;&\lambda_1=g,\quad \lambda_2=g+2,\\
(0,\pi),(\pi,0):\;&\lambda_{1,2}=\pm\sqrt{g(g+2)}.
\end{aligned}
\end{equation}
Therefore, $g=-2$ and $g=0$ act as special points separating five regimes: $g<-2$, $g=-2$, $-2<g<0$, $g=0$, and $g>0$. Figure~\ref{fig:point} shows representative flow lines in the $(x,y)$ plane for selected $g$.

\paragraph{(1) Strong negative coupling: $g<-2$.}
For $g<-2$, $(\pi,\pi)$ has $\lambda_1<0$ and $\lambda_2<0$ and is a stable node (sink), while $(0,0)$ is an unstable node (source). The remaining fixed points $(0,\pi)$ and $(\pi,0)$ satisfy $\lambda_1\lambda_2<0$ and are saddles. Accordingly, all trajectories converge to $(\pi,\pi)$.

\paragraph{(2) Special point I: $g=-2$.}
At $g=-2$, Eq.~\eqref{eq:BAB_xy} becomes
\begin{equation}
\dot{x}=\sin x-\sin y,\qquad
\dot{y}=-\sin x+\sin y ,
\end{equation}
so that $\dot{x}+\dot{y}=0$ and hence
\begin{equation}
x+y=u=\mathrm{const}.
\end{equation}
Trajectories are confined to straight lines $y=-x+u$. Equilibria satisfy $\sin x=\sin y$, giving two families of equilibrium manifolds,
\begin{equation}
(\mathrm{A}):\ x-y=2k\pi,\qquad
(\mathrm{B}):\ x+y=(2m+1)\pi ,
\end{equation}
with $k,m\in\mathbb{Z}$. The Jacobian has eigenvalues $\lambda_1=0$ and $\lambda_2=\cos x+\cos y$.

Along family~(A), $\cos y=\cos x$ and thus $\lambda_2=2\cos x$, so $x-y=2k\pi$ decomposes into transversely stable segments ($\cos x<0$) and unstable segments ($\cos x>0$). Along family~(B), $\cos y=-\cos x$, giving $\lambda_2=0$; therefore $x+y=(2m+1)\pi$ is neutrally stable.

In terms of the original angles, equilibria correspond to phase-locked motion with
\begin{equation}
\dot{\theta}_1=\dot{\theta}_2=\dot{\theta}_3=2\sin x^{*}.
\end{equation}
Except when $\sin x^*=0$, the molecule approaches a phase-locked rotating state: the relative angles remain fixed while all particles co-rotate with the same constant angular velocity.

\paragraph{(3) Special point II: $g=0$.}
The dynamics at $g=0$ are analogous. The equilibrium manifolds are
\begin{equation}
(\mathrm{A}):\ x+y=2k\pi,\qquad
(\mathrm{B}):\ x-y=(2m+1)\pi ,
\end{equation}
and the Jacobian eigenvalues are $\lambda_1=0$ and $\lambda_2=-(\cos x+\cos y)$.
Along family~(A), $\lambda_2=-2\cos x$, so $x+y=2k\pi$ contains transversely stable segments ($\cos x>0$) and unstable segments ($\cos x<0$). Along family~(B), $\lambda_2=0$ and the manifold is neutral.

In the original variables, $g=0$ implies
\begin{equation}
\dot{\theta}_1=0,\qquad
\dot{\theta}_3=0,\qquad
\dot{\theta}_2=\sin(\theta_1-\theta_2)+\sin(\theta_3-\theta_2),
\end{equation}
so $\theta_1$ and $\theta_3$ remain fixed while $\theta_2$ relaxes until
$\sin(\theta_1-\theta_2)+\sin(\theta_3-\theta_2)=0$.

\paragraph{(4) Positive coupling: $g>0$.}
For $g>0$, $(0,0)$ is a sink, $(\pi,\pi)$ is a source, and $(0,\pi)$ and $(\pi,0)$ are saddles; hence all trajectories converge to $(0,0)$.

\paragraph{(5) Oscillatory regime: $-2<g<0$.}
For $-2<g<0$, $(0,0)$ and $(\pi,\pi)$ are saddles, while $(0,\pi)$ and $(\pi,0)$ have purely imaginary eigenvalues
\begin{equation}
\lambda_{1,2}=\pm i\sqrt{-g(g+2)},
\end{equation}
and are therefore centers. By symmetry, it suffices to analyze $(x^*,y^*)=(0,\pi)$. Linearization gives
\begin{equation}
\dot{\delta x}=-(1+g)\delta x+\delta y,\qquad
\dot{\delta y}=-\delta x+(1+g)\delta y,
\end{equation}
which implies
\begin{equation}
\ddot{\delta x}+ \omega^2 \delta x=0,
\qquad
\omega=\sqrt{-g(g+2)} .
\end{equation}
A convenient parametrization is
\begin{equation}
\delta x(t)=a\cos(\omega t),
\end{equation}
with amplitude $a$ set by the initial condition. Using
$\delta y=\dot{\delta x}+(1+g)\delta x$, one obtains
\begin{equation}
\delta y(t)=a\big[(1+g)\cos(\omega t)-\omega\sin(\omega t)\big]
          =a\cos(\omega t+\phi_0),
\end{equation}
where
\begin{equation}
\cos\phi_0=1+g,\qquad
\sin\phi_0=\omega.
\end{equation}
Thus the small-amplitude motion takes the form
\begin{equation}
\begin{pmatrix}
x(t) \\[4pt]
y(t)
\end{pmatrix}
=
\begin{pmatrix}
x^{*} \\[4pt]
y^{*}
\end{pmatrix}
+
a
\begin{pmatrix}
\cos(\omega t) \\[4pt]
\cos(\omega t+\phi_0)
\end{pmatrix},
\label{eq:nonreciprocal}
\end{equation}
corresponding to closed orbits around the center with a phase lag determined solely by $g$.

Finally, we numerically integrate Eq.~\eqref{eq:BAB_xy} for representative $g$ values and obtain the flow lines shown in Fig.~\ref{fig:point}, in agreement with the stability analysis above.

\supsection{Dependence of molecular dynamics on nonreciprocal coupling $g$}

In the main text, we focused on the case of nonreciprocal coupling $g=-1$. More generally, the molecular dynamics can be \emph{continuously} tuned by varying the coupling strength $g$. To illustrate this dependence, we consider a single [B--A--B] molecule at fixed noise amplitudes, $D_{\mathrm{r}} = 10^{-3}$ and $D_{\mathrm{t}} = D_{\mathrm{r}}/3$, identical to those used in Fig.~1(d) and Fig.~3 of the main text, and systematically vary $g$.

\begin{figure}
    \centering
    \includegraphics[width=0.9\linewidth]{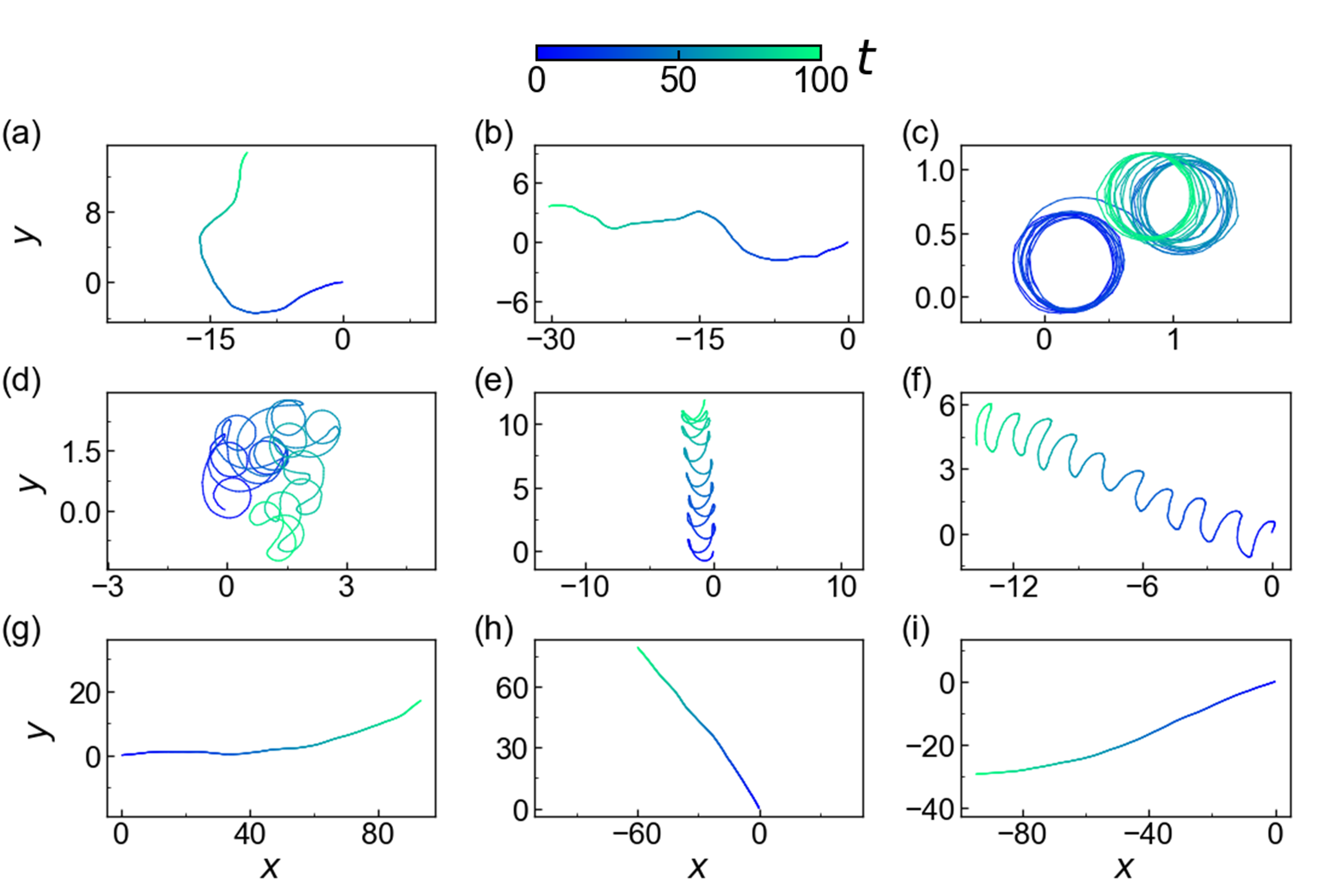}
    \caption{Trajectories of the center of mass of a [B-A-B] molecule for different values of $g$. Panels (a)–(i) correspond to $g=-3.0,\,-2.5,\,-2.0,\,-1.6,\,-1.0,\,-0.4,\,0,\,0.5,$ and $1.0$, respectively. Trajectories are color-coded by time as indicated by the color bar at the top.
}
    \label{fig:gtraj}
\end{figure}

Figure~\ref{fig:gtraj} shows the resulting trajectories of the molecular center of mass for several representative values of $g$. For $g<-2$ and $g>0$, the trajectories are locally straight, with only a slowly varying propulsion direction. In contrast, for intermediate values $-2<g<0$, the molecule exhibits a characteristic wriggling forward motion. As $g$ increases from $-2$ toward $0$, the persistence time of the propulsion direction increases, and the trajectories become progressively smoother.

\begin{figure}
    \centering
    \includegraphics[width=0.9\linewidth]{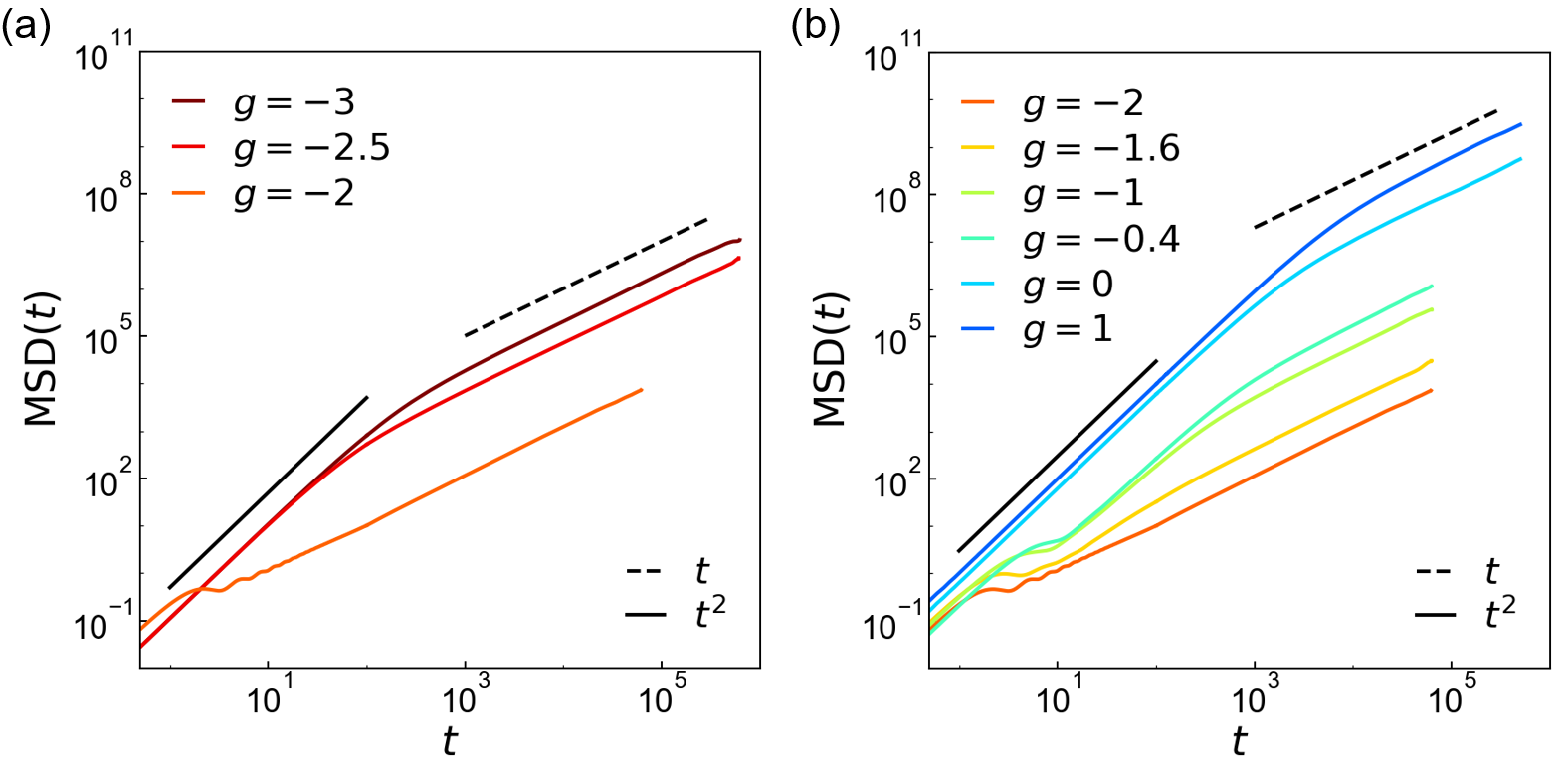}
    \caption{Mean-squared displacement $\mathrm{MSD}(t)$ of the center of mass of a [B--A--B] molecule for different values of $g$. Solid and dashed guide lines indicate the reference scalings $\mathrm{MSD}(t)\sim t^{2}$ (ballistic) and $\mathrm{MSD}(t)\sim t$ (diffusive), respectively.
}
    \label{fig:gMSD}
\end{figure}

The corresponding mean-squared displacement (MSD) of the molecular center of mass is shown in Fig.~\ref{fig:gMSD}. For reference, we indicate the ballistic scaling $\mathrm{MSD}(t)\sim t^{2}$ (solid line) and the diffusive scaling $\mathrm{MSD}(t)\sim t$ (dashed line). For $g<-2$ and $g>0$, the MSD exhibits a clear crossover from ballistic to diffusive behavior, consistent with a slowly decorrelating propulsion direction. In the intermediate regime $-2<g<0$, the MSD depends on $g$ more intricately and attains its minimum near the special point $g=-2$. This trend is consistent with the noise-free analysis in Sec.~\ref{sec:no_noise}. For $-2<g<0$, the deterministic dynamics produce wriggling motion, whereas at $g=-2$ the relative angles phase-lock while all three particles rotate synchronously with a common constant angular velocity. As a result, the center-of-mass executes uniform circular motion. In the presence of noise, the orbit center undergoes slow diffusion, so the MSD remains finite but is suppressed compared to nearby values of $g$.

\supsection{Mean-squared displacement and net-force autocorrelation for different molecular codings}

\supsubsection{Numerical evaluation}

To characterize translational motion, we compute the mean-squared displacement (MSD) of the molecular center of mass, defined in Eq.~(5) of the main text. We evaluate it as a time average over reference times $t_0$ along a steady-state trajectory,
\begin{equation}
\mathrm{MSD}(t)
\equiv
\left\langle
\left|
\mathbf{R}(t_{0}+t)-\mathbf{R}(t_{0})
\right|^{2}
\right\rangle_{t_{0}} =
\frac{1}{T-t}\int_{t_{\mathrm{ini}}}^{t_{\mathrm{ini}}+T-t}
\left|
\mathbf{R}(t_{0}+t)-\mathbf{R}(t_{0})
\right|^{2}\,dt_{0},
\label{eq:msd_int}
\end{equation}
where $t_{\mathrm{ini}}$ marks the start of the steady-state segment after discarding the initial transient, and $T$ is the total duration of the averaging window. For the [A--A--A], [A--B--B], and [A--A--B] codings, we set $T=1\times10^{6}$ and compute time averages by sampling reference times at intervals of $\Delta t_{0}=10^{-2}$. For the [B--A--B] coding, the rapid orientational fluctuations necessitate a smaller numerical time step; accordingly, we use $T=2\times10^{5}$ and take $\Delta t_{0}=10^{-3}$. We further perform an ensemble average over 100 independent trajectories. These settings are used to generate the MSD(t) curves shown in Fig.~3(a) of the main text.



To rationalize the MSD behavior, we compute the autocorrelation function of the net active propulsion force defined in Eq.~(6) of the main text,
\begin{equation}
C_{\mathbf{F}_{\mathrm{net}}}(t)
=
\left\langle 
\mathbf{F}_{\mathrm{net}}(t_{0}+t)
\cdot
\mathbf{F}_{\mathrm{net}}(t_{0})
\right\rangle_{t_{0}},
\label{eq:corre}
\end{equation}
which we evaluate as the finite-time average
\begin{equation}
C_{\mathbf{F}_{\mathrm{net}}}(t)
=
\frac{1}{T-t}\int_{t_{\mathrm{ini}}}^{t_{\mathrm{ini}}+T-t}
\mathbf{F}_{\mathrm{net}}(t_{0}+t)\cdot
\mathbf{F}_{\mathrm{net}}(t_{0})\,dt_{0}.
\label{eq:corre_int}
\end{equation}
Here $\mathbf{F}_{\mathrm{net}}(t)=\sum_{i=1}^{3}F_{0}\mathbf{n}_i(t)$ is the instantaneous net propulsion force of a molecule, with $\mathbf{n}_i(t)=(\cos\theta_i(t),\sin\theta_i(t))$. For [A--A--A], [A--B--B], and [A--A--B], $C_{\mathbf{F}{\mathrm{net}}}(t)$ decays monotonically; we therefore use an averaging window $T=10^{6}$ and sample reference times at intervals of $\Delta t_{0}=10^{-2}$. For [B--A--B], $C_{\mathbf{F}{\mathrm{net}}}(t)$ exhibits pronounced oscillations with a slowly drifting period, so averaging over overly long trajectories would smear out this structure. We thus use a shorter window, $T=2\times10^{2}$, to resolve the intrinsic oscillations and take $\Delta t_{0}=10^{-3}$. These settings are used to compute the $C_{\mathbf{F}_{\mathrm{net}}}(t)$ curves shown in Fig.~3(b) of the main text.


\supsubsection{Behavior of the net propulsion force $\mathbf{F}_{\mathrm{net}}(t)$}

In this section, we analyze the autocorrelation of the net propulsion force for different three-particle molecular sequences. The instantaneous net propulsion force is
\begin{equation}
\mathbf F_{\mathrm{net}}(t)
=
\sum_{i=1}^3 F_0\,\mathbf n_i(t)
\equiv
F_{\mathrm{eff}}\,\mathbf N(t),
\end{equation}
where $\mathbf n_i(t)=(\cos\theta_i(t),\sin\theta_i(t))$ is the propulsion direction of particle $i$, and $\mathbf N(t)=(\cos\Theta(t),\sin\Theta(t))$ defines the instantaneous molecular orientation.

We first consider the sequences \textbf{[A--A--A]}, \textbf{[A--B--B]}, and \textbf{[A--A--B]}. In the deterministic limit, their internal orientational dynamics can be analyzed in the same $(x,y)$ representation used for \textbf{[B--A--B]} by defining $x \equiv \theta_1-\theta_2$ and $y \equiv \theta_3-\theta_2$. One finds that the stable fixed point is $(x^*,y^*)=(0,0)$ for \textbf{[A--A--A]} and \textbf{[A--B--B]}, while for \textbf{[A--A--B]} there are two symmetry-related stable fixed points at $(x^*,y^*)=(0,\pi)$ and $(\pi,0)$.

\begin{figure}[t]
    \centering
    \includegraphics[width=0.8\linewidth]{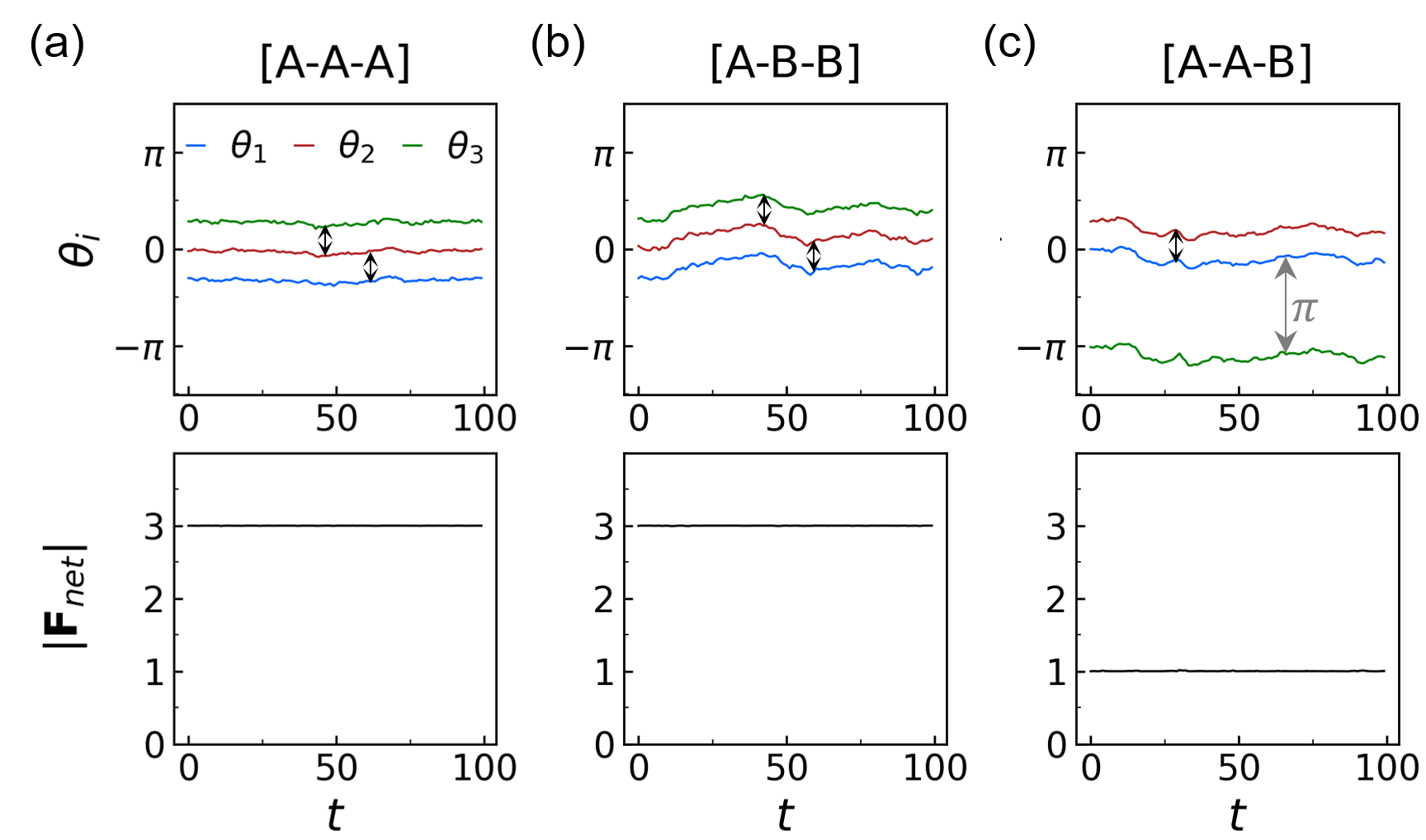}
    \caption{
        Evolution of the individual orientations $\theta_i(t)$ (upper panels) and the magnitude of the net propulsion force $|\mathbf{F}_{\mathrm{net}}(t)|$ (lower panels) for the sequences (a) \textbf{[A--A--A]}, (b) \textbf{[A--B--B]}, and (c) \textbf{[A--A--B]}. For clarity, the $\theta_i(t)$ traces in the upper panels are vertically offset by $0.3\pi$ (black arrows) when they would otherwise overlap.
        }
    \label{fig:FnetAAA}
\end{figure}

With thermal noise present, after a short transient these molecules remain localized near the corresponding stable internal configurations. Figures~\ref{fig:FnetAAA}(a)--(c) show representative time traces of $\theta_i(t)$ together with the magnitude $|\mathbf F_{\mathrm{net}}(t)|$. For clarity, the $\theta_i(t)$ traces are vertically shifted by $0.3\pi$ when they would otherwise overlap. Aside from small fluctuations and a slow drift of the global phase, the internal angles remain phase-locked. Specifically, for \textbf{[A--A--A]} and \textbf{[A--B--B]} the configuration is approximately $\phi_0(t)+[0,0,0]$, while for \textbf{[A--A--B]} it is $\phi_0(t)+[0,0,\pi]$, where $\phi_0(t)$ denotes a global phase. These locked configurations determine the net-force magnitude: in the \textbf{[A--A--A]} and \textbf{[A--B--B]} cases the three propulsion directions add constructively, giving $|\mathbf F_{\mathrm{net}}|\simeq 3F_0$, whereas for \textbf{[A--A--B]} one particle is approximately out of phase with the other two, yielding $|\mathbf F_{\mathrm{net}}|\simeq F_0$.

Because the internal configuration is stable while the overall molecular orientation undergoes rotational diffusion, the direction of $\mathbf F_{\mathrm{net}}$ performs a random walk. As a result, the autocorrelation
\begin{equation}
C_{\mathbf{F}_{\mathrm{net}}}(t)
=
\left\langle
\mathbf F_{\mathrm{net}}(t_0+t)\cdot
\mathbf F_{\mathrm{net}}(t_0)
\right\rangle_{t_0}
\end{equation}
is expected to exhibit an approximately exponential decay for these three sequences, consistent with Fig.~3(b) of the main text. This behavior is well described by the standard active Brownian particle form:
\begin{equation}
\begin{aligned}
C_{\mathbf{F}_{\mathrm{net}}}(t)
&=
F_{\mathrm{eff}}^{\,2}
\left\langle
\mathbf N(t_0+t)\cdot \mathbf N(t_0)
\right\rangle_{t_0} \\
&=
F_{\mathrm{eff}}^{\,2}
\left\langle
\cos\!\big[\Theta(t_0+t)-\Theta(t_0)\big]
\right\rangle_{t_0}.
\end{aligned}
\end{equation}
Let $\Delta\Theta(t)\equiv\Theta(t_0+t)-\Theta(t_0)$. Under rotational diffusion, $\Delta\Theta(t)$ is a zero-mean Gaussian variable with
\begin{equation}
\langle \Delta\Theta(t)^2\rangle = 2D_{\mathrm{eff}}\,t .
\end{equation}
Using $\langle \cos X\rangle=\mathrm{Re}\langle e^{iX}\rangle$ and $\langle e^{iX}\rangle=\exp(-\langle X^2\rangle/2)$ for a zero-mean Gaussian $X$, we obtain
\begin{equation}
\left\langle \cos\Delta\Theta(t)\right\rangle
=
\exp(-D_{\mathrm{eff}}\,t).
\end{equation}
and therefore
\begin{equation}
C_{\mathbf{F}_{\mathrm{net}}}(t)
=
F_{\mathrm{eff}}^{\,2}\,e^{-D_{\mathrm{eff}} t},
\label{eq:C_exp}
\end{equation}
which is the usual autocorrelation of an active Brownian particle with effective propulsion $F_{\mathrm{eff}}$ and effective rotational diffusion coefficient $D_{\mathrm{eff}}$.

For the \textbf{[B--A--B]} sequence, the internal angles $\theta_i(t)$ exhibit sustained oscillations reminiscent of the noise-free dynamics [see Fig.~2(d) in the main text], with noise-induced fluctuations in both amplitude and frequency. Consequently, $|\mathbf F_{\mathrm{net}}(t)|$ also oscillates in time, and $C_{\mathbf{F}_{\mathrm{net}}}(t)$ develops pronounced oscillations, as shown in Fig.~3(b) of the main text.

\supsubsection{Derivation of $\mathrm{MSD}(t)$ from $C_{\mathbf{F}_{\mathrm{net}}}(t)$ }

The overdamped translational dynamics of particle $i$ is
\begin{equation}
\begin{aligned}
\dot{\mathbf{r}}_i(t)
&= \frac{1}{\gamma} \!\left[
F_0\,\mathbf{n}_i(t)
- \nabla_i \!\left(
\sum_{j\in\mathcal B(i)} U_{\mathrm{spr}}(r_{ij})
+ \sum_{k (\neq i)} U_{\mathrm{WCA}}(r_{ik}) 
\right)
\right]
+ \sqrt{2 D_\text{t}}\,\bm{\eta}^t_i(t),
\end{aligned}
\label{eq:trans_particle}
\end{equation}
where the translational noises $\bm{\eta}^t_i(t)$ are Gaussian white noises with zero mean and unit variance,
\begin{equation}
\langle \bm{\eta}^\mathrm{t}_i(t) \rangle = \mathbf{0},\qquad
\langle \bm{\eta}^\mathrm{t}_i(t)\bm{\eta}^\mathrm{t}_j(t')\rangle
= \delta_{ij}\delta(t-t') \mathbf{1}.
\end{equation}
For a three-particle molecule, the center-of-mass position is
\begin{equation}
\mathbf r_{\mathrm{CM}}(t) = \frac{1}{3}\sum_{i=1}^3 \mathbf r_i(t). \end{equation}
Summing Eq.~\eqref{eq:trans_particle} over $i$ and using $\sum_i\nabla_i(\cdots)=0$ for internal pair interactions, we obtain
\begin{equation}
\dot{\mathbf r}_{\mathrm{CM}}(t)
=
\frac{1}{3\gamma}\mathbf F_{\mathrm{net}}(t)
+\sqrt{2 D_\text{t}}\,\bm{\eta}^{\mathrm t}_{\mathrm{CM}}(t),
\label{eq:trans_com_raw}
\end{equation}
where the net active force is
\(
\mathbf F_{\mathrm{net}}(t)=\sum_{i=1}^3 F_0\,\mathbf n_i(t)
\), and
\begin{equation}
\bm{\eta}_{\mathrm{CM}}^{\mathrm t}(t)=\frac{1}{3}\sum_{i=1}^3 \bm{\eta}^t_i(t).
\end{equation}
It follows immediately that
\begin{equation}
\langle \bm{\eta}_{\mathrm{CM}}^{\mathrm t}(t) \rangle = \mathbf{0} ,\qquad
\langle \bm{\eta}_{\mathrm{CM}}^{\mathrm t}(t)\bm{\eta}_{\mathrm{CM}}^{\mathrm t}(t') \rangle
= \frac{1}{3}\,\delta(t-t')\,\mathbf{1}.
\label{eq:eta_cm_corr}
\end{equation}
The center-of-mass displacement over a lag time $t$ is
\begin{equation}
\Delta\mathbf r(t)_{\mathrm{CM}}\equiv \mathbf r_{\mathrm{CM}}(t_0+t)-\mathbf r_{\mathrm{CM}}(t_0)
=\int_0^t \dot{\mathbf r}_{\mathrm{CM}}(t_0+s)\,ds .
\end{equation}
Assuming a steady state with time-translation invariance, the MSD can be written as
\begin{align}
\mathrm{MSD}(t)
&\equiv \left\langle\big|\Delta\mathbf r_{\mathrm{CM}}(t)\big|^2\right\rangle_{t_0}
= \left\langle
\int_0^t ds \int_0^t ds'\,
\dot{\mathbf r}_{\mathrm{CM}}(t_0+s)\cdot\dot{\mathbf r}_{\mathrm{CM}}(t_0+s')
\right\rangle_{t_0}.
\label{eq:msd_cv}
\end{align}
Using time-translation invariance,
\begin{equation}
\langle \mathbf F_{\mathrm{net}}(t_0+s)\cdot \mathbf F_{\mathrm{net}}(t_0+s')\rangle_{t_0}
=
C_{\mathbf{F}_{\mathrm{net}}}(|s-s'|), 
\end{equation}
together with the standard identity
\begin{equation}
\int_0^t ds\int_0^t ds'\,f(|s-s'|)
=2\int_0^t (t-\tau)\,f(\tau)\,d\tau,
\label{eq:conv_identity}
\end{equation}
the active-force contribution becomes
\begin{equation}
\label{eq:msd_active_term}
\frac{1}{9\gamma^2}\int_0^t ds\int_0^t ds'\,
\left\langle
\mathbf F_{\mathrm{net}}(t_0+s)\cdot \mathbf F_{\mathrm{net}}(t_0+s')
\right\rangle_{t_0}
=
\frac{2}{9\gamma^2}\int_0^t (t-\tau)\,C_{\mathbf{F}_{\mathrm{net}}}(\tau)\,d\tau.
\end{equation}

For the noise contribution, Eq.~\eqref{eq:eta_cm_corr} implies
\begin{equation} 
\left\langle \bm{\eta}^{\mathrm t}_{\mathrm{CM}}(t_0+s)\cdot \bm{\eta}_{\mathrm{CM}}^{\mathrm t}(t_0+s') \right\rangle_{t_0} = \frac{d}{3}\delta(s-s'),
\end{equation}
where $d$ is the spatial dimension ($d=2$ here). Therefore,
\begin{equation}
\label{eq:msd_noise_term}
2D_{\mathrm t}\int_0^t ds\int_0^t ds'\,
\left\langle
\bm{\eta}^{\mathrm t}(t_0+s)\cdot \bm{\eta}^{\mathrm t}(t_0+s')
\right\rangle_{t_0}
=
\frac{2dD_{\mathrm t}}{3}\,t,
\end{equation}
Combining Eqs.~\eqref{eq:msd_active_term} and \eqref{eq:msd_noise_term}, we obtain
\begin{equation}
\mathrm{MSD}(t)
=
\frac{2}{9\gamma^2}\int_0^t (t-\tau)\,
C_{\mathbf{F}_{\mathrm{net}}}(\tau)\,d\tau
+\frac{4D_{\mathrm t}}{3}\,t.
\label{eq:msd_c}
\end{equation}
For the sequences \textbf{[A--A--A]}, \textbf{[A--B--B]}, and \textbf{[A--A--B]}, substituting Eq.~\eqref{eq:C_exp} into Eq.~\eqref{eq:msd_c} yields
\begin{equation}
\mathrm{MSD}(t)
=
\frac{4D_{\mathrm t}}{3}\,t
+\frac{2}{9\gamma^2}F_{\mathrm{eff}}^2
\left[
\frac{t}{D_{\mathrm{eff}}}
-\frac{1-e^{-D_{\mathrm{eff}}t}}{D_{\mathrm{eff}}^{2}}
\right],
\end{equation}
which corresponds to Eq.~(7) in the main text.

\supsection{Determination of the steady-state cluster fraction in MIPS}

For the motility-induced phase separation (MIPS) simulations shown in Fig.~4 of the main text, the system contains $N=14{,}400$ molecules when each molecule consists of three particles (i.e., $43{,}200$ particles in total). For monomeric systems, we use $N=43{,}200$ single-particle molecules. All simulations are performed in a square periodic box at fixed particle number density $\rho=0.6$.

Clusters are identified from a distance-based connectivity criterion at the particle level: two particles are considered connected if their separation satisfies $r_{ij}<1.10\sigma$ \cite{Fily_2012_Athermala}
. A cluster of molecules is then defined as the set of molecules whose constituent particles belong to the same connected particle network (for monomers this reduces to the usual particle-based definition). The size of the largest cluster, $N_{\mathrm{lc}}(t)$, is therefore reported in units of molecules, and the largest-cluster fraction is defined as
\begin{equation}
\alpha(t)=\frac{N_{\mathrm{lc}}(t)}{N}.
\end{equation}

\begin{figure}[htbp]
\centering
\includegraphics[width=0.9\linewidth]{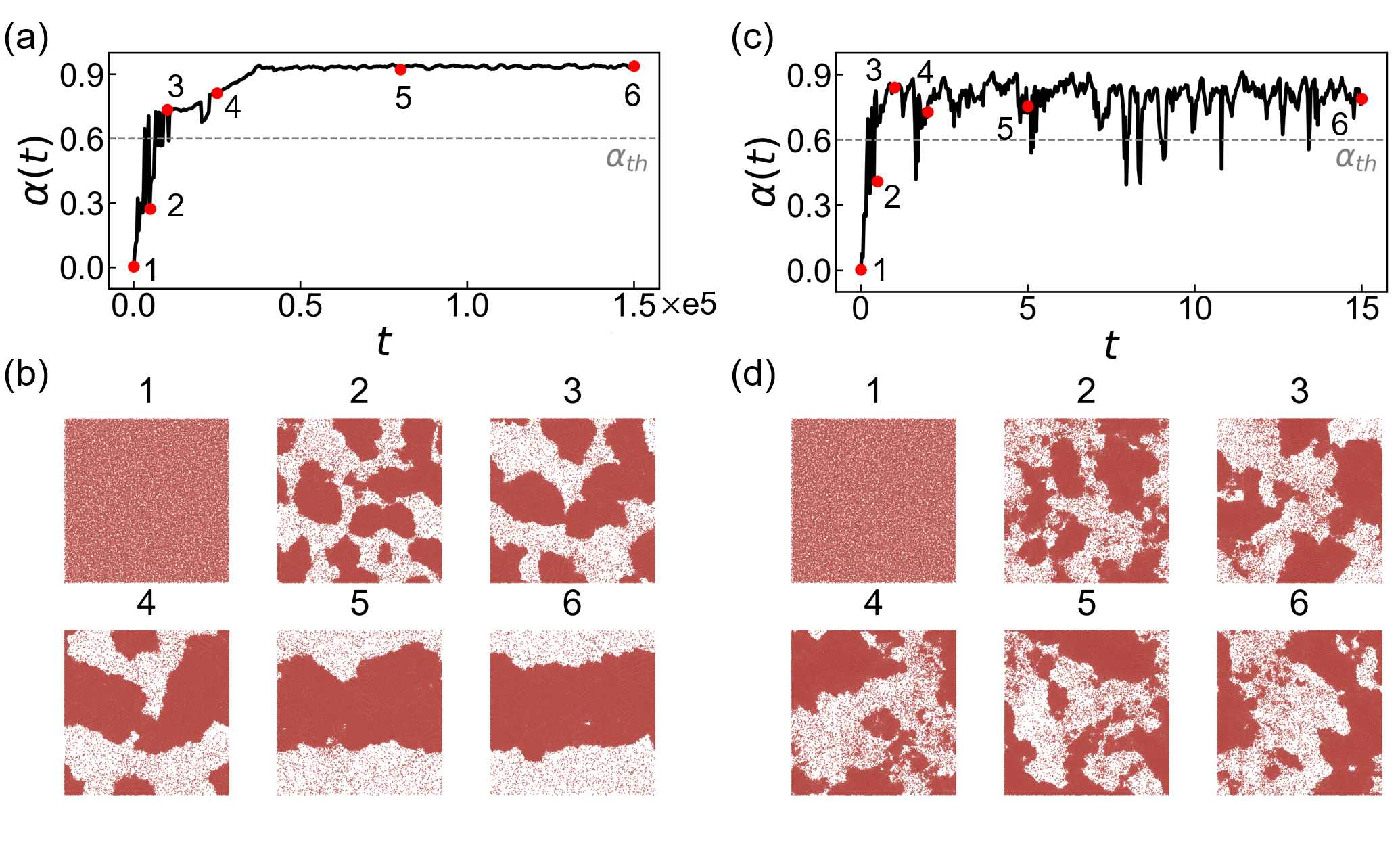} 
\caption{ 
(a,c) Time evolution of the largest-cluster fraction $\alpha(t)$ for the [A] monomer at $F_0=0.1$ (a) and $F_0=1000$ (c).
(b,d) Representative snapshots at the times marked by the red symbols in panels (a) and (c), respectively.
}
\label{fig:mipst}
\end{figure}

Figure~\ref{fig:mipst} shows $\alpha(t)$ for the [A] monomer at two representative propulsion strengths, $F_0=0.1$ [Fig.~\ref{fig:mipst}(a)] and $F_0=1000$ [Fig.~\ref{fig:mipst}(c)]. In both cases, $\alpha(t)$ grows from near zero and approaches a plateau, indicating the emergence of a quasi-steady cluster fraction. The plateau value is smaller and the temporal fluctuations are substantially larger at $F_0=1000$ than at $F_0=0.1$, consistent with the more intermittent cluster dynamics at high activity. Representative snapshots at the times marked in Fig.~\ref{fig:mipst}(a,c) are shown in Fig.~\ref{fig:mipst}(b,d).

Because both the steady-state value and the fluctuation amplitude of $\alpha(t)$ depend on $F_0$, we determine the steady-state cluster fraction $\bar{\alpha}$ using an adaptive protocol:

\paragraph{(i) Decorrelation time.}
We compute the normalized autocorrelation function of the order parameter,
\begin{equation}
C_{\alpha}(\tau)=
\frac{\langle \alpha(t+\tau)\,\alpha(t)\rangle_t}{\langle \alpha(t)^2\rangle_t},
\qquad C_{\alpha}(0)=1,
\end{equation}
and define the decorrelation time $\tau_c$ by $C_{\alpha}(\tau_c)=1/e$. We then sample $\alpha(t)$ at intervals $\tau_c$ to obtain approximately independent measurements.

\paragraph{(ii) Stationarity test and relaxation fraction.}
To ensure that the trajectory has reached a statistically stationary regime, we discard an initial fraction $\chi$ of the time series and split the remaining data into two consecutive windows of equal duration. Denoting the corresponding window averages by $\bar{\alpha}_1$ and $\bar{\alpha}_2$, we perform a Welch two-sample $t$-test for the null hypothesis $\bar{\alpha}_1=\bar{\alpha}_2$ at significance level $\delta=0.05$. If the test yields $p>\delta$, we regard $\alpha(t)$ as stationary over the retained interval. For each parameter set, $\chi$ is chosen as the minimal discarded fraction that satisfies this criterion.

\paragraph{(iii) Production averaging.}
For each value of $F_0$, we run the simulation long enough to discard the initial fraction $\chi$, and then continue for an additional time $2\times10^{4}/F_0$ during which $\alpha(t)$ is sampled every $\tau_c$. The steady-state cluster fraction $\bar{\alpha}$ is obtained by averaging these samples.

\paragraph{(iv) Independent runs.}
For each $F_0$, we perform multiple independent simulations and report $\bar{\alpha}$ as the average over independent realizations.

\supsection{MIPS for different molecular codings}

\begin{figure}[h]
    \centering
    \includegraphics[width=0.9\linewidth]{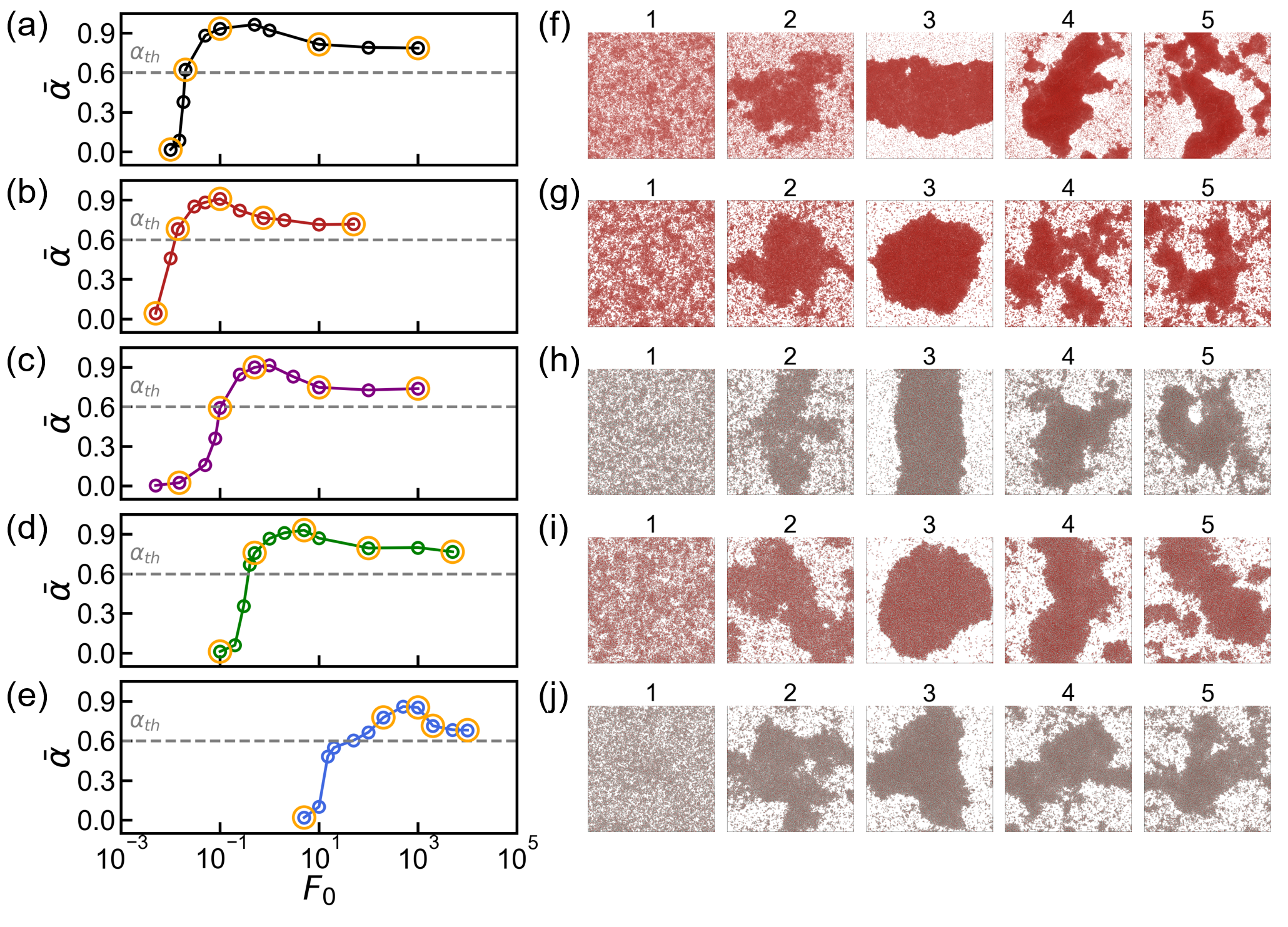}
       \caption{
Motility-induced phase separation (MIPS) for different molecular codings.
(a--e) Replot of Fig.~4(a) in the main text, shown in separate panels. Steady-state largest-cluster fraction $\bar{\alpha}$ as a function of the propulsion strength $F_0$ for each coding. The gray dashed line indicates the threshold $\alpha_{\mathrm{th}}=0.6$.
(f--j) Representative snapshots corresponding to the orange open circles in panels (a--e).
}
\label{fig:selfcorree3}
\end{figure}

Figure~\ref{fig:selfcorree3}(a--d) shows the steady-state largest-cluster fraction $\bar{\alpha}$ as a function of the propulsion strength $F_0$ for different molecular codings. Simulation parameters are identical to those used in Fig.~4 of the main text. Representative configuration snapshots corresponding to selected points are shown in Fig.~\ref{fig:selfcorree3}(e--h).

\clearpage
\bibliography{supplement}